# First observation of electron emission from the DD threshold resonance

K. Czerski[1]*, R. Dubey[1], M. Kaczmarski[1], A. Kowalska[2], N. Targosz-Sleczka[1], G. Das Haridas[1], M. Valat[1]

[1] *Institute of Physics, University of Szczecin, 70-451 Szczecin, Poland*
[2] *Physics Department, Maritime University of Szczecin, 70-500 Szczecin, Poland*

konrad.czerski@usz.edu.pl

Electron emission in the deuteron-deuteron reaction supporting existence of the single-particle threshold resonance in $^4$He has been observed for the first time. The measured electron energy spectrum and the electron-proton branching ratio agree very well with the assumed electron-positron pair creation decay of the $0^+$ resonance state to the ground state and the detailed Monte Carlo simulations of the experimental energy spectrum.

**Introduction**

The $^4$He level structure at excitation energies below 30 MeV seems to be very well known for last decades and could be successfully applied for description of nuclear reaction by means of the multichannel R-matrix parametrization [1]. Recent ab-initio structure calculations of the four-nucleon system applying realistic nucleon-nucleon interactions and the microscopic cluster approach [2-4] confirmed the known level structure of $^4$He, as well. Therefore, it was very surprising that the last precise measurements of the $^2$H(d,p)$^3$H reaction cross-section performed on the deuterated Zr target at deuteron energies down to 5 keV [5] pointed to a strong contribution of a $0^+$ resonance placed close to the DD reaction threshold. This new state is supposed to be a single-particle resonance having the 2+2 cluster structure, which results in very small and energy dependent resonance width. From the theoretical point of view [5,6], the existence of the $0^+$ threshold resonance arises from a weak coupling between the states of the 1+3 and 2+2 clustering (see Fig. 1). Whereas the s- and p- wave resonances below and close to the DD threshold at the excitation energy of about 24 MeV have almost the pure 1+3 structure, the known negative parity levels ($0^-$, $1^-$, $2^-$) at the excitation energy of about 28 MeV show predominantly the 2+2 clustering of the relative angular momentum L=1. Thus, in analogy to the first excited state of $^4$He with $J^\pi=0^+$ that can be interpreted as the s-wave resonance of the 1+3 cluster [16], a similar s-wave state should be also expected for the 2+2 cluster nearby the DD threshold [6].

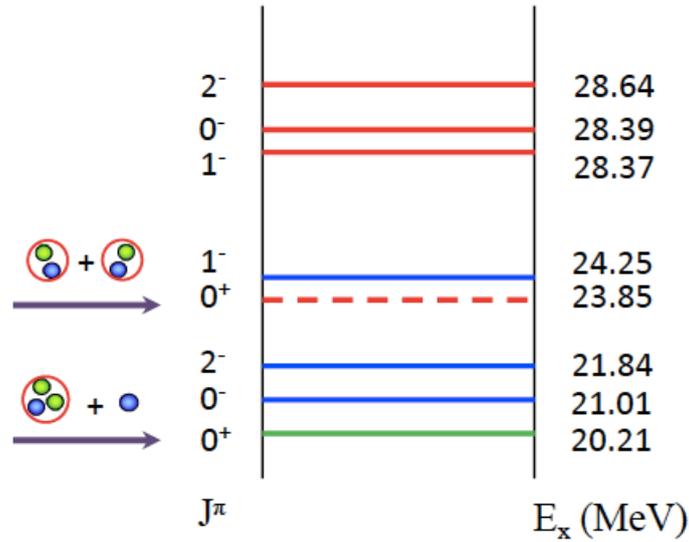

Fig. 1. Schematic energy levels of $^4$He representing s- and p-wave resonances for the 1+3 (blue) and 2+2 (red) cluster states. The arrows show the corresponding thresholds which are close to the 0$^+$ resonances. The DD threshold resonance is marked with the red dashed line.

According to the theoretical calculations based on the E0 energy weighted sum rule [6] the DD threshold resonance should have a large partial width for the electron-positron pair creation and therefore, the detection of emitted electrons with a continuous energy spectrum up to 23 MeV in the DD reactions could provide an additional argument for the existence of the threshold resonance. The latter would be of large importance for nuclear reaction rates of astrophysical plasmas and commercial applications of the DD fusion reactions. The contribution of the $e^+e^-$ channel to the DD reaction cross section should increase for decreasing deuteron energies far below Coulomb barrier of about 350 keV, where the DD reactions are also enhanced by the electron screening effect. The charges of reacting nuclei are shielded by surrounding electrons of the medium leading to reduction of the Coulomb barrier height and an increase of the penetration factor through the barrier. This effect is especially important for the dense stellar plasma of giant planets, brown and white dwarfs [7], where the reaction rates can increase by many orders of magnitude [8]. Investigation of the low-energy nuclear reactions in metallic environments with their quasi-free electrons provides a unique occasion to compare experimental results with theoretical predictions. Many experiments carried out in the past showed large discrepancies for the screening energy values obtained for both metallic [9] and gaseous [10] targets. Thus, a precise determination of the threshold resonance parameters could help us to understand the nature of the observed enhancement of the DD reaction cross-section in the energy region, where the electron screening as well as excitation of the threshold resonance take place [5].

In this letter, we report the first observation of the electron emission in the low-energy DD reactions which can originate from the internal pair creation decay of the threshold resonance. The experimental results will be compared to the theoretical calculations of the reaction branching ratio, showing that the $e^+e^-$ transition to the ground state of $^4$He should be the strongest reaction channel at very low deuteron energies. The experimental analysis will be also supported by careful Monte Carlo simulations using the Geant 4 code [11].

**Experimental setup**

The experiment was performed at the eLBRUS Ultra High Vacuum Accelerator Facility of the Szczecin University in Szczecin, Poland [12]. A deuterium beam was accelerated to energies ranging between 6 and 16 keV, with the constant current beam of 40 µA, using the magnetically analyzed single-charged atomic and molecular deuterium ions. The beam was impinged on a 0.5 mm thick $ZrD_2$ target that was tilted at 45° to the beam, resulting in the beam spot size of 7x12 mm. To reduce the systematic uncertainties, only one EG ORTEC silicon detector of the 1 mm thickness and 100 mm$^2$ detection area, situated at the backward angle 135° was used for all charged particles emitted: protons, tritons and $^3$He particles as well as electrons and positrons produced by the DD reactions. Utilizing a single detector enabled to reduce its distance to the target down to 6 cm and increase its solid angle, which was necessary due to the rapidly dropping reaction cross section for lowering deuteron energies. The traditional analog NIM bin system was used to process the energy signal created in the detector, and data were acquired via the TUKAN MCA.

In the front of the detector, an Al absorption foil was placed for two different reasons. First, the detector should be protected against elastically scattered beam deuterons. Secondly, the foil thickness was set to 3.7 µm to also absorb 0.8 MeV $^3$He particles and reduce the energy of the emitted 1.02 MeV tritons below 0.6 MeV in order to detect electrons free of background up to 1 MeV, which is the highest electron energy detectable by our Si detector (see Fig. 2). For channels above 300 corresponding to energies higher than 1.5 MeV, a small number of 3 MeV protons scattered in the absorption foil could be observed additionally to the natural background.

To be sure that the continuum part of the charged particle spectrum results from electron/positron emission, we also applied a 2 mm thick Si detector that fully absorb electrons of maximum energy 2 MeV. Additionally, the absorption Al foil in the front of the detector was of 45 μm thickness, which was enough to stop emitted tritons and reduce the energy of protons to 1.8 MeV, but practically did not change the energy spectrum of high energy electrons/positrons. The experimental spectrum for this case is presented in Fig. 3. A prominent bump at energy 0.75 MeV would correspond to the average energy loss of high energy electron/positrons in the detector.

The energy calibration of the detector was performed with a $^{241}$Am alpha-source and $^{22}$Na, $^{60}$Co, $^{204}$Tl beta sources with and without the Al absorption foil. In Fig. 4, the electron energy spectrum of $^{204}$Tl decaying by two different β- transitions of the Q= 345 keV and 763 keV is presented.

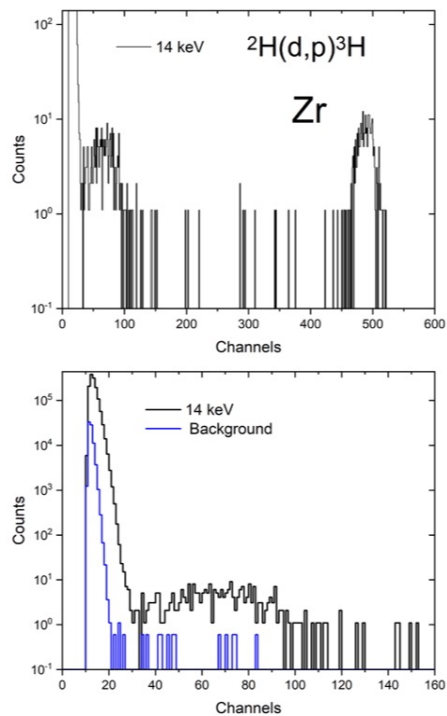

Fig. 2. Experimental energy spectrum measured at the deuteron energy 14 keV using the 1 mm thick Si detector and 3.7 μm thick Al absorption foil. The lower part is extended and shows the experimental background.

**Simulation of the electron energy spectrum**

As explained before, we expect that the threshold resonance in $^4$He at the excitation energy of 23.84 MeV should decay by the internal electron-positron pair creation. To calculate the final energy spectrum of electrons registered in the Si detector, we performed a series of the Monte Carlo simulations using the Geant 4 code [11]. The computational ability can be demonstrated in Fig. 4, where the electron spectrum of the $^{204}$Tl radioactive source could be described very precisely. The increase of the counting rate at very low energies corresponds to the low energy photons produced due to the bremsstrahlung and secondary recombination effects, e.g. about 80 keV X-rays induced in the gold surface layer of the detector (see also Fig. 5). The Geant 4 calculations were performed for different thicknesses of detectors and absorption foils and all projectile energies. The final energy spectra are the result of 100 simulation runs with $5 \times 10^7$ electrons and $5 \times 10^7$ photons each, leading to $5 \times 10^9$ incident events in the detector.

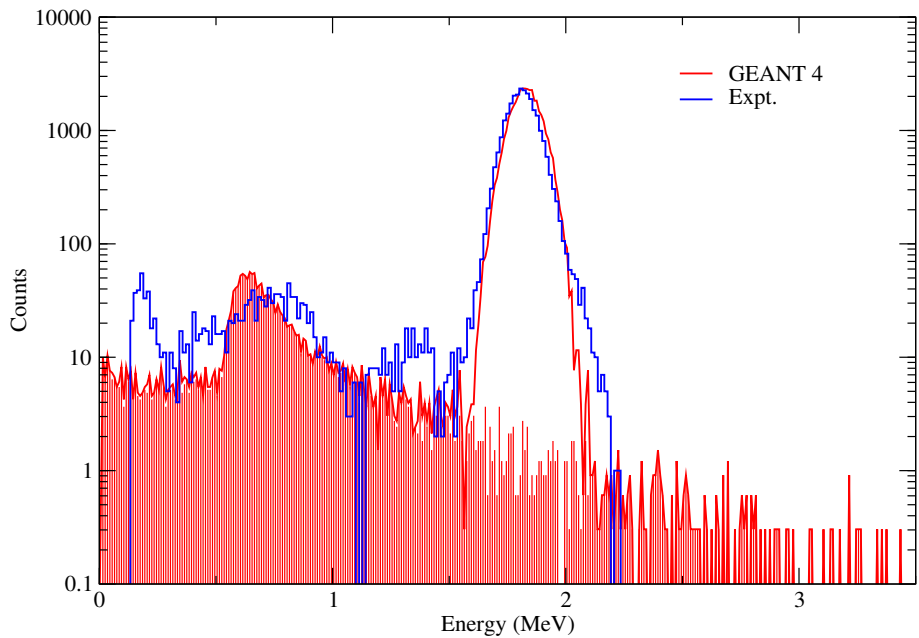

Fig. 3. Experimental energy spectrum measured at the deuteron energy 20 keV using the 2 mm thick Si detector and 45 μm thick absorption Al foil (in blue). The Geant 4 calculated spectrum is presented in red.

The electron/positron energy spectrum calculated for the threshold resonance decay as it is expected for the thin (1mm) detector is depicted in Fig. 5 – the original energy distribution of the internal pair creation is inserted in the same figure. The maximum energy of emitted electrons amounts to 22.73 MeV, but only electrons with energy of 1 MeV or less can be fully absorbed in the detector. Therefore, the high energy electrons of the pair creation for which the detector is transparent will be registered at the energies lower than 1 MeV. The electron spectrum calculated using the Geant 4 simulations was compared to the experimental ones in Fig. 5, showing that the average energy loss of electrons/positrons in the detector is about 450 keV. The theoretical spectra were normalized to the experimental counting rate observed in the energy region above the triton line, i.e. 0.6-1.0 MeV. In such a way, we can determine the percentage of the measured electrons to the total number of emitted due to the threshold resonance decay. This factor was assumed to be constant for all projectile energies and equal to 0.42±0.02. The stopping power and ranges of positrons in Si are only about 2 % percent lower than for electrons [17]. It means that the energy spectrum of detected positrons does not need to be simulated separately and the number of emitted electrons can be simply increased by a factor of two.

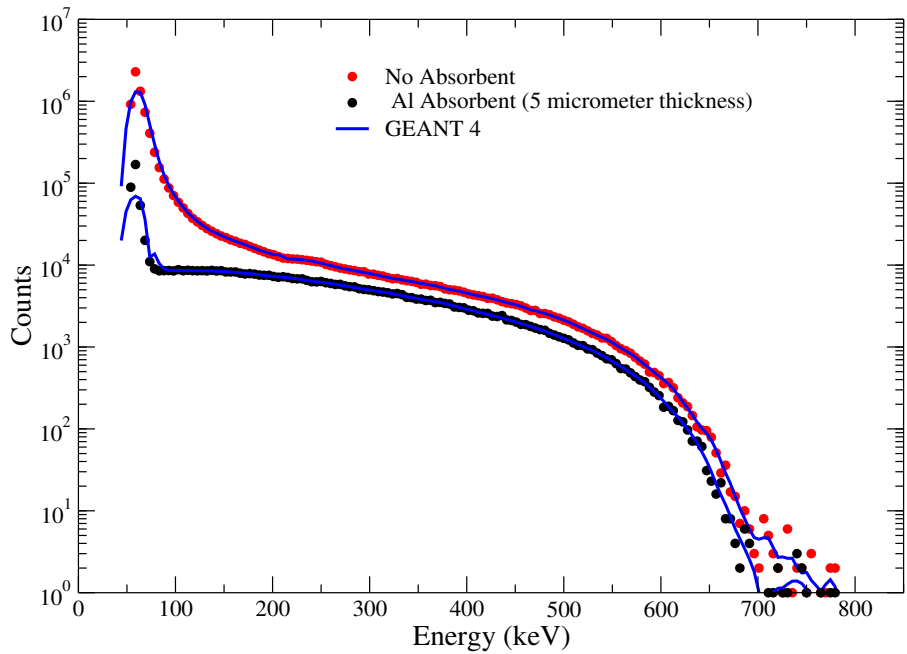

Fig. 4. Comparison between experimental and simulated energy spectra of the β- radioactive source $^{204}$Tl. The spectra were measured with and without the 5 μm Al absorber placed at the front of the detector.

The calculated energy spectrum shows a strong increase of the counting rate for energies below 150 keV due to absorbed photons similar to the spectra measured for the $^{204}$Tl radioactive source. Without the photonic contributions, the energy spectrum strongly drops. The theoretical spectrum fits very well the experimental data obtained for different projectile energies.

In the case of the thicker (2 mm) Si detector and thicker absorption foil, the experimental spectrum can also be described quite well (see Fig. 3). The average energy loss of electrons/positrons was predicted for energies a little lower than experimentally observed, which can probably be explained by a large uncertainty of stopping power values of high energy electrons and positrons. Procedure for determining the electron/proton branching ratio was similar to that applied for the thin detector measurements. Experimental results obtained for other combinations of the detector and absorption foil thicknesses will be presented in a separated work [18]. All this is consistent with the emission of high-energy electrons.

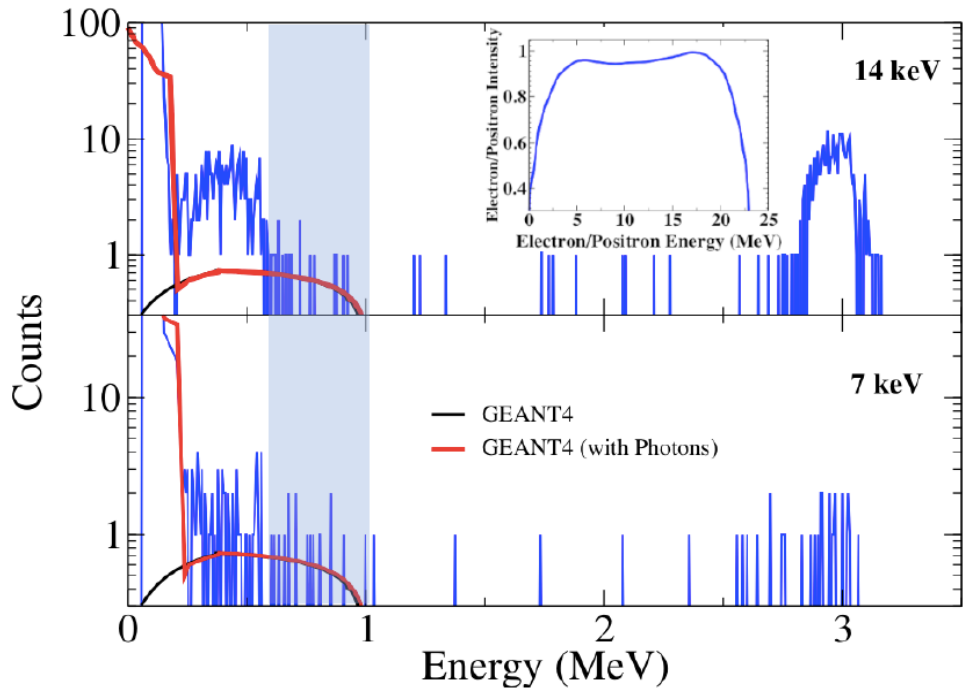

Fig. 5. Comparison between the charged particle spectra measured at two different deuteron energies and the calculated electron spectrum. Upper panel: The two distinct peaks are protons, and tritons along with electrons/positrons from the DD reaction at the deuteron energy 14 keV. Bottom panel: the spectrum obtained for 7 keV. Solid lines correspond to the Geant 4 simulated spectra: electronic part only (black line) and electrons together with low energy photons (red line). The insert in the upper part presents the energy spectrum of emitted electrons/positrons from the threshold resonance according to the Geant 4 simulation. The shaded box corresponds to the electron/positron counting integration area.

**Theoretical analysis**

As shown previously [6], the threshold DD resonance in ⁴He should have a large partial width for the electron-positron pair creation. The enhancement of the reaction yield measured for the proton channel at the deuteron energies down to 6 keV [5] could be explained by both the electron screening effect and the destructive interference of the threshold resonance with the known broad resonances of ⁴He. Thus, determination of the electron-to-proton branching ratio could be an independent prove for existence of the DD threshold resonance. The experimentally determined values are presented in Fig. 5. Both reaction channels are studied by the same detector, therefore the error bars arise only from the statistical uncertainties.

The screened nuclear reaction cross section for the ²H(d,p)³H reaction can be parameterized as follows [13]:

$$\sigma_{scr}(E) = \frac{1}{\sqrt{E(E+U_e)}} S(E) \exp\left(-\sqrt{\frac{E_G}{E+U_e}}\right) = \frac{1}{\sqrt{EE_G}} P(E+U_e) S(E) \quad (1)$$

where $S(E)$ is the astrophysical S-factor, and the s-wave penetration factor through the Coulomb barrier $P(E)$ is given by

$$P(E) = \sqrt{\frac{E_G}{E}} \exp\left(-\sqrt{\frac{E_G}{E}}\right) \quad (2)$$

Here, $E$ and $E_G$ stay for the center-mass energy and the Gamow energy equal to $E_G = \frac{2\pi^2}{137^2} \mu c^2$, respectively, and $\mu$ is the reduced mass. The screening energy $U_e$ determines reduction of the Coulomb barrier height. For the deuteron energy below 50 keV, the astrophysical S-factor is very well known and can be presented as a linear energy function [14].

The cross section of the 0⁺ threshold resonance can be simply expressed by the Breit-Wigner formula:

$$\sigma_{res} = \frac{\pi}{k^2} \frac{\Gamma_d \Gamma_p}{(E-E_{res})^2 + \frac{1}{4}\Gamma_{tot}^2} \quad (3)$$

where the deuteron partial width strongly depends on energy and is given by:

$$\Gamma_d(E) = 2k\, a\, P(E) \frac{\hbar^2}{\mu a^2} |\theta_d|^2 \quad (4)$$

Here $k$ and $a$ denote for the deuteron wave number and the channel radius, respectively. $\theta_d$ is the reduced resonance width equal to unity, assuming the single-particle resonance structure. For the deuteron energy studied, the total resonance width $\Gamma_{tot}$ is dominated by the deuteron partial width and the other contributions can be neglected. Additionally, taking into account that both the resonance energy $E_{res}$ and the total resonance width is much smaller than the deuteron energy, the resonance cross-section expression can be further simplified:

$$\sigma_{res} \cong \frac{\pi}{k^2} \frac{2k\, a\, P(E+U_e)\frac{\hbar^2}{\mu a^2}\Gamma_p}{E^2} = \frac{\pi}{k} P(E+U_e) \frac{2\hbar^2}{\mu a} \frac{\Gamma_p}{E^2} \tag{5}$$

Consequently, the total cross section of the $^2$H(d,p)$^3$H reaction can be presented as a sum of two contributions: the structureless 'flat' component describing the known broad and overlapping resonances in $^4$He and the component owning to the narrow threshold resonance. In the case of the incoherent sum, we get:

$$\sigma_p = \frac{\pi}{k} P(E+U_e) \left[ \frac{k}{\pi} \frac{1}{\sqrt{EE_G}} S(E) + \frac{2\hbar^2}{\mu a} \frac{\Gamma_p}{E^2} \right] \tag{6}$$

As shown previously [5], the 0$^+$ threshold resonance can, however, interfere with the known broad 0$^+$ resonances of $^4$He contributing to the excitation function $\sigma_{flat}$. Thus, the coherent 0$^+$ contribution to the total cross-section can be expressed as follows:

$$\sigma_p^{0+} = \sigma_{flat}^{0+} + \sigma_{res} + 2\sqrt{\sigma_{flat}^{0+}\, \sigma_{res}} \cos(\varphi_{flat}^{0+} - \varphi_{res}) \tag{7}$$

where $\varphi_{res}$ is the resonance phase shift given by:

$$\text{tg}\, \varphi_{res} = \frac{\Gamma_{tot}}{2(E-E_{res})} \to 0 \tag{8}$$

takes very low values and can be neglected in Eq. (7). $\varphi_{flat}^{0+}$ represents the nuclear phase shift of the $\alpha_0=\langle ^1S_0|0^+|^1S_0\rangle$ transition matrix element [15]. Therefore, in the coherent case, the proton emission cross section reads as follows:

$$\sigma_p = \frac{\pi}{k} P(E+U_e) \left[ \frac{k}{\pi} \frac{1}{\sqrt{EE_G}} S(E) + \frac{2\hbar^2}{\mu a} \frac{\Gamma_p}{E^2} + \left( \frac{k}{\pi} \frac{S(E)}{3} \frac{1}{\sqrt{EE_G}} \cdot \frac{2\hbar^2}{\mu a} \frac{\Gamma_p}{E^2} \right)^{1/2} \cos(\varphi_{flat}^{0+}) \right] \tag{9}$$

In the interference term of the expression above, we take into account that the $\alpha_0=\langle ^1S_0|0^+|^1S_0\rangle$ transition makes about 1/3 of the total cross section [15].

For the electron-positron pair creation, we assume that only the threshold resonance contributes. Contribution of the known broad resonances should be much smaller due to higher angular momenta of the corresponding transitions, and the resonance strength would be spread over a large energy range. Therefore, the corresponding cross section takes the resonant form:

$$\sigma_{res} = \frac{\pi}{k} P(E + U_e) \frac{2\hbar^2}{\mu a} \frac{\Gamma_{pair}}{E^2} \qquad (10)$$

Finally, it is clear that the resulting branching ratio between the pair creation and proton emission cross-sections does not depend on the penetration factor and the screening energy, anymore.

In Fig. 6, the experimental data of the branching ratio is fitted by the theoretical curves taking into account or neglecting the interference effect. Only free fitting parameter was the partial resonance width of the pair creation. The cross-section function of the proton emission was taken from the work [5], where the proton width was determined to be $\Gamma_p$=40 meV and the nuclear phase shift $\varphi_{flat}^{0+}$=115°. We can see that the interference effect does not strongly influence the energy dependence of the branching ratio in the investigated deuteron energy region. Larger differences can be observed only at the lower energies. For a comparison, the theoretical curve for the proton resonance width of 20 meV is also given. Once again, the differences can be found at the lower energies. The partial resonance widths of the pair creation estimated for different cases take similar values to those of the proton channel.

Table 1. Ratio of the proton and pair creation partial resonance widths estimated for different fitting cases as described in the text.

|  | $\Gamma_p = 40\ meV$ | | $\Gamma_p = 20\ meV$ |
| --- | --- | --- | --- |
|  | coherent | incoherent | coherent |
| $\Gamma_{pair}\ (meV)$ | 55±7 | 71±12 | 55±7 |
| $\Gamma_{pair}/\Gamma_p$ | 1.4±0.2 | 1.8±0.3 | 2.8±0.4 |

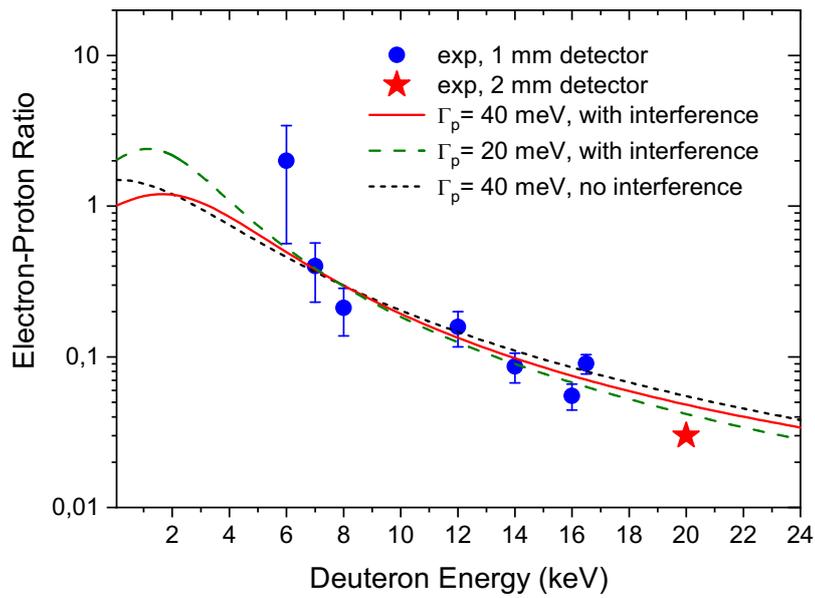

Fig. 6. Experimental electron-proton branching ratio estimated for the electron energy loss 0.6-1.0 MeV using the thin (1 mm) Si detector and compared with the theoretical calculations. The high energy point for the deuteron energy 20 keV measured with the thick (2mm) Si detector was adjusted to the thin detector measurements.

**Discussion and conclusions**

We have observed for the first time the electron emission in the DD reactions at very low energies which can be attributed to the the decay of the threshold resonance by means of the internal pair creation. This $0^+$ resonance was previously observed in the $^2$H(d,p)$^3$H reaction preceding in both metallic Zr and gaseous environments [5,6]. The energy spectrum of the measured electrons/positrons agrees very well with the results of simulations performed using the Geant 4 code. Additionally, the energy dependence of the electron-proton branching ratio determined for the deuteron energies between 6 and 16 keV can be also very well explained with excitation of the threshold resonance. For broad resonances of $^4$He, we would expect a constant branching ratio not exceeding the lowest measured value of about 0.03. This confirms that their small contribution could be neglected in our first analysis. The fit curves depicted in Fig. 6 describe two different models for taking into account the threshold resonance: with and without the interference effect (Eq. 6 and Eq. 9). According to the fitting procedure, a constant contribution to the branching ratio should be smaller than 0.2. The differences between the theoretical curves are very small in the studied energy range. Similarly, the change of the proton partial width which was estimated in the earlier study of the $^2$H(d,p)$^3$H reaction within the experimental uncertainty does not influence the shape of the theoretical curve significantly. Likewise, the different theoretical models lead to only slightly different values of the partial pair-creation widths. The largest one equal to 71±12 meV was obtained for the incoherent threshold resonance amplitude. All the values are within the range predicted on the basis of the E0 energy weighted sum rule [6]. To distinguish between the different resonance parameters, measurements at the deuteron energies below 5 keV will be necessary, which will be increasingly difficult due to dropping cross-section. Theoretically, the $0^+$ threshold resonance can decay by the internal electron conversion, as well. But the process is many orders of magnitude less probable than pair creation at such a high excitation energy and would result in a sharp, discrete line in the electron spectrum, which is not observed. Similarly, a decay to other excited states in $^4$He is strongly suppressed because of weak coupling between 2+2 and 3+1 cluster states. The suppression factor should be of order $10^{-7}$ which corresponds to the ratio of partial resonance widths of the proton and deuteron channels observed in the $^2$H(d,p)$^3$H reaction.

Despite the experimental difficulties, this work provides a strong and independent arguments for the existence of the $0^+$ threshold resonance in the DD reactions and might have large consequences for the nuclear astrophysics and the nuclear fusion applied studies.

**Acknowledgments**

This project has received funding from the European Union's Horizon 2020 research and innovation programme under grant agreement No 951974.

**References**

[1]   D. R. Tilley, H.R. Weller, G.M. Hale, Nucl. Phys. A **541**, 1 (1992).

[2]   H. M. Hoffmann, G.M. Hale, Phys. Rev. C **77**, 044002 (2008).

[3]   K. Arai et al., Phys. Rev. Lett. **107**, 132502 (2011).

[4]   S. Aoyama *et al.*, Few-Body Syst **52,** 97 (2012).

[5]   K. Czerski, et al., EPL **113**, 22001 (2016).

[6]   K. Czerski, Phys. Rev. C (Letters) **106**, L011601 (2022).

[7]   E. E. Salpeter, Aust. J. Phys. **7**, 373 (1954).

[8]   S. Ichimaru, H. Kitamura, Physics of Plasmas **6**, 2649 (1999).

[9]   A. Huke et al., Phys. Rev. C **78**, 015803 (2008).

[10]  C. Spitaleri et al., Phys. Lett. B **755**, 275 (2016).

[11]  S. Agostinelli et al., Nucl. Instrum.Meth. A **506**, 250 (2003).

[12]  K. Kaczmarski et al., Acta Phys. Pol. B **45**, 509 (2014).

[13]  K. Czerski et al., Europhys. Lett. **68**, 363 (2004).

[14]  R.E. Brown, N. Jarmie, Phys. Rev. C **41**, 1391 (1990).

[15]  H. Paetz gen. Schieck, Eur. Phys. J. A **44**, 321 (2010).

[16]  E. Hiyama, B.F. Gibson, M. Kamimura, Phys. Rev. C **70**, 031001(R) (2004).

[17]  M. J. Berger and S. M. Seltzer, report NBSIR 82-2550 (1982).

[18]  G. Das Haridas, R. Dubey, K. Czerski, M. Kaczmarski, A. Kowalska, N. Targosz-Sleczka, M. Valat, to be published.